\documentclass{article}

\usepackage{arxiv}

\usepackage[utf8]{inputenc} 
\usepackage[T1]{fontenc}    
\usepackage{hyperref}       
\usepackage{url}            
\usepackage{booktabs}       
\usepackage{amsfonts}       
\usepackage{nicefrac}       
\usepackage{microtype}      
\usepackage{lipsum}

\title{HLA predictions from the bronchoalveolar lavage fluid samples of five patients at the early stage of the Wuhan seafood market COVID-19 outbreak}

\author{
   Ren\'e L.~Warren\\
  Genome Sciences Centre, BC Cancer\\
  Vancouver, BC, V5Z 4S6, Canada \\
  \texttt{rwarren@bcgsc.ca} \\
   \And
 Inan\c{c} ~Birol \\
  Genome Sciences Centre, BC Cancer\\
  Vancouver, BC, V5Z 4S6, Canada \\
  \texttt{ibirol@bcgsc.ca} \\
}

\begin{document}
\maketitle

\begin{abstract}
We are in the midst of a global viral pandemic, one with no cure and a high mortality rate. The Human Leukocyte Antigen (HLA) gene complex plays a critical role in host immunity. We predicted HLA class I and II alleles from the transcriptome sequencing data prepared from the bronchoalveolar lavage fluid samples of five patients at the early stage of the COVID-19 outbreak. We identified the HLA-I allele A*24:02 in four out of five patients, which is higher than the expected frequency (17.2\%) in the South Han Chinese population. The difference is statistically significant with a p-value less than $10^{-4}$. Our analysis results may help provide future insights on disease susceptibility.

\textbf{Availability:} The tool used to perform the reported analysis results, HLAminer, is available from \url{https://github.com/bcgsc/hlaminer}. Predictions are available for download from \url{https://www.bcgsc.ca/downloads/btl/SARS-CoV-2/BAL}
\end{abstract}

\keywords{COVID-19 \and SARS-Cov-2 \and Human Leukocyte Antigen (HLA) \and RNA-Seq \and bronchoalveolar  lavage  fluid \and HLA-A*24:02 \and HLA-DPA1*02:02 \and HLA-DPB1*05:01}

\section{Introduction}
SARS-CoV-2 infections have reached global pandemic proportions in early 2020, affecting over 2M people worldwide (as of this writing) and showing no signs of easing, except in a few jurisdictions where strict quarantine measures were implemented early on. The resulting coronavirus disease (COVID-19) has a relatively high (${\sim}$3.4\%) mortality rate \cite{rajgor} – a figure that varies widely between jurisdictions due to factors yet to be determined. Currently, no vaccines or effective treatments are available. Most current data analysis efforts are, understandably, focused on the virus itself for the purpose of vaccine development and tracking its evolution for diagnostics and infection monitoring purposes.\par
Curiously, it is estimated that as high as 18-30\% or more of the population may be asymptomatic to SARS-CoV-2 infections \cite{mizumoto} \cite{nishiura}, while other affected individuals exhibit mild to severe to critical symptoms of infection. Thus, gaining insights on host susceptibility to the coronavirus is clearly another important aspect that needs to be worked on and understood.
One would expect a link between host immunity genes and susceptibility or resistance to infection. The Human Leukocyte Antigen (HLA) gene complex includes two classes of such genes, which encode the Major Histocompatibility Complex (MHC). Proteins of the MHC present (class I) internally- or (class II) externally-derived antigenic determinants (epitopes) to T cells, which upon recognition of the epitope-complex, will mount an immune response to defend against viral and bacterial infections. HLA genes are therefore cornerstone to acquired immunity in humans. HLA alleles have also been shown to be factors in susceptibility or resistance to certain diseases, and their frequency and composition in human populations vary widely (http://allelefrequencies.net). A previous study found HLA class I genes HLA-B*46:01 and HLA-B*54:01 to be associated with the 2003 severe acute respiratory syndrome (SARS) coronavirus infections in Taiwan \cite{lin} – a related disease to the current pandemic.\par
For over a decade, high-throughput transcriptome sequencing (RNA-Seq) has proven a worthy instrument for measuring changes of gene expression in human diseases and beyond \cite{wang}. Transcriptome analysis has the potential to reveal key genes that are modulated in response to infections, but also to reveal the HLA composition of affected individuals. A few years ago, our group developed an approach for mining high-throughput next-generation shotgun sequencing data for the purpose of HLA determination \cite{warren}, which has since been applied in a broader clinical context \cite{brown}. \par
Here, we report our initial observations based on transcriptome sequencing libraries prepared from the bronchoalveolar lavage fluid samples of five patients at the early stage of the Wuhan seafood market pneumonia coronavirus outbreak (see Methods). We identified the HLA class I allele A*24:02 and class II haplotype DPA1*02:02-DPB1*05:01 in four out of five individuals. Although HLA-A*24:02 is common in some populations, the prevalence observed (80\%) is higher than the allele frequency in the Chinese population (17.2\%) — the presumed ethnicity of the patients in the reported study.\par

\section{Methods}
We downloaded MGISEQ-2000RS paired-end (150 bp) RNA-Seq reads from libraries prepared from the bronchoalveolar lavage fluid samples of five patients (\url{https://www.ebi.ac.uk/ena/data/view/PRJNA605983} Accessions: SRX7730880-SRX7730884 denoted in the tables below as Patient 1-5, respectively). We note that these are metagenomics RNA samples, and, although not explicitly noted, we think that they were prepared for the primary purpose of identifying and characterizing the novel coronavirus at the outbreak epicentre. On each dataset, we ran HLAminer \cite{warren} in targeted assembly mode with default values (v1.4; contig length $\geq$200bp, seq. identity $\geq$99\%, score $\geq$1000), predicting HLA class I (HLA-I) and class II (HLA-II) alleles and report 4-digit (HLA allele/protein) resolution when top-scoring predictions are unambiguous. Otherwise the 2-digit (allele group) resolution is reported.\par

\begin{table}[h!]
\hrule \vspace{0.1cm}
\caption{\label{tab:hlaI}HLA-I predictions from the bronchoalveolar lavage fluid samples of five patients at the early stage of the Wuhan seafood market pneumonia coronavirus outbreak. Highest-scoring \textit{HLAminer} predictions are shown for HLA class I genes A, B and C. Common HLA alleles between two or more patients are highlighted in bold face.}
\centering
\begin{tabular}{ccccc}
\toprule
\textbf{Patient 1} & \textbf{Patient 2} & \textbf{Patient 3} & \textbf{Patient 4} & \textbf{Patient 5}\\
\midrule
 A*01:01P &      A*30:01P &      \textbf{A*24:02P} &     \textbf{A*24:02P} &             A*29 \\
 \textbf{A*24:02P}  &    A*02:06P &      A*26:01P &      -       &       \textbf{A*24} \\\hline
 B*35/B*57 &     \textbf{B*51:01P} &     B*15:01P &      B*40:01P &              B*54:01P\\ 
 B*48 &  B*13:02P &      \textbf{B*51:01P} &     B*13:01P &              B*07:05P \\\hline
 C*08:72P &      C*14:02P &      C*15:02P &      C*04:03P &              C*15 \\
 \textbf{C*06:02P} &     \textbf{C*06:02P} &     C*03:03P &      C*03:04P &              - \\
\bottomrule
\end{tabular}
\end{table}

\begin{table}[ht]
\hrule \vspace{0.1cm}
\caption{\label{tab:hlaII}HLA-II predictions from the bronchoalveolar lavage fluid samples of five patients at the early stage of the Wuhan seafood market pneumonia coronavirus outbreak. Highest-scoring \textit{HLAminer} predictions are shown for HLA class II genes DPA1, DPB1, DQA1, DQB1, DRA, DRB1, DRB3, DRB4 and DRB5. Missing class II genes or (-) denote the absence of predictions. Common HLA alleles between two or more patients are highlighted in bold face.}
\centering
\begin{tabular}{ccccc}
\toprule
\textbf{Patient 1} & \textbf{Patient 2} & \textbf{Patient 3} & \textbf{Patient 4} & \textbf{Patient 5}\\
\midrule
\textbf{DPA1*02:02P} &	DPA1*02:01P &	\textbf{DPA1*02:02P} &	\textbf{DPA1*02:02P} &	DPA1*04:01P\\
-	& \textbf{DPA1*02:02P} &	-	& -	& - \\\hline
\textbf{DPB1*05:01P} &	DPB1*13:01P &	\textbf{DPB1*05:01P} &	\textbf{DPB1*05:01P} &	- \\
DPB1*02:02P &	\textbf{DPB1*05:01P} &	- &	- &	- \\\hline
DQA1*01:03P &	DQA1*01:02P &	DQA1*03:01P &	DQA1*01:01P &	- \\
DQA1*05	& DQA1*02:01P	& -	& DQA1*05:01P &	- \\\hline
\textbf{DQB1*03:01P} &	DQB1*05:02P &	DQB1*04:01P &	\textbf{DQB1*03:01P} &	- \\
DQB1*06:01P &	DQB1*02:01P &	DQB1*03:02P &	-	& - \\\hline
\textbf{DRA*01:01P} &	\textbf{DRA*01:01P} &	\textbf{DRA*01:01P} &	-	& - \\
- &	- &	- &	- &	- \\\hline
DRB1*08:03P &	DRB1*07:01P &	DRB1*04:03P &	DRB1*15:02P &	DRB1*04 \\
-	& DRB1*16:02P & 	DRB1*04:05P & 	DRB1*11:01P	& - \\\hline
-	& - &	- &	DRB3*02:02P	& - \\
- &	- &	- &	- &	- \\\hline
- &	\textbf{DRB4*01:01P} &	\textbf{DRB4*01:01P} &	- &	- \\
- &	- &	- &	-	& - \\\hline
- &	DRB5*02:02P &	- &	DRB5*01:02 &	- \\
- &	-	& - &	- &	- \\
\bottomrule
\end{tabular}
\end{table}

\section{Results}
We predicted and compiled the likely HLA class I (Table 1) and II (Table 2) alleles for five patients at the early stage of the COVID-19 outbreak in Wuhan, China. Although the bronchoalveolar lavage (BAL) fluid samples were initially utilized to identify and characterize the novel coronavirus (similar justification in \cite{wu}), BAL metagenomics samples are expected to contain host cells / nucleic acids (DNA/RNA). Because HLA genes are expressed at the surface of all human nucleated cells, RNA-Seq data can be employed to determine HLA profiles from BAL samples.\par
We observe the HLA-A*24:02 allele in four out of five (80\%) patients (Table 1). HLA-A*24 is a common group of alleles in Southeastern Asian populations, and the frequency of HLA-A*24:02 allele is high, especially in indigenous Taiwanese populations, reaching as high as 86.3\%. However, our understanding is that the five patients are from the Wuhan market area, and the associated A*24:02 frequency in Chinese population is typically 17.2\%, a value that is statistically significantly less than our observed frequency of 80\% (p$<10^{-4}$, 1-sided z-test). Also of note, the HLA class II DPA1*02:02 and DPB1*05:01 haplotype predicted in patients 1 to 4 (80\%), genes DQB1*03:01 in patients 1 and 4, and DRB4*01:01 in patients 2 and 3 (Table 2).\par

\section{Discussion and Conclusions}
HLA-A*24 has not been previously reported as a risk factor for SARS infection \cite{sun}, but there are reports of disease association with HLA-A*24:02, notably with diabetes \cite{adamashvili} \cite{noble} \cite{nakanishi} \cite{kronenberg}, which is a reported potential risk factor in COVID-19 patients \cite{guan}. Both DPA1*02:02 and DPB1*05:01 occur at relative high frequency (44.8\% and 31.3\%, n=1490) in Han Chinese \cite{chu}, and associations of those HLA type II alleles with narcolepsy \cite{ollila} and Graves' disease \cite{chu}, both autoimmune disorders, have been reported in that population. Further, a GWAS study found a link between HLA-DPB1*05:01 and chronic hepatitis B in Asians, and it has been suggested that this risk allele may impact one's ability to clear viral infections \cite{ollila} \cite{kamatani}.\par
HLA also informs vaccine development. This knowledge would help prioritize SARS-CoV-2 derived epitopes predicted to be stable HLA binders \cite{nguyen} \cite{prachar}. Future studies into host susceptibility and resistance to SARS-CoV-2 infections are sorely needed as they may help us better manage and mitigate risks of infections. We chose to communicate our early findings in this domain to facilitate rapid development of response strategies.\par

\subsection*{Grant information}
This work was supported by Genome BC and Genome Canada [281ANV]; and the National Institutes of Health [2R01HG007182-04A1]. The content of this paper is solely the responsibility of the authors, and does not necessarily represent the official views of the National Institutes of Health or other funding organizations.

\bibliographystyle{unsrt}  


\begin{thebibliography}{15}
%
\bibitem{rajgor}
Rajgor, D.D. et al. (2020) The many estimates of the COVID-19 case fatality rate. The Lancet. doi:10.1016/S1473-3099(20)30244-9

\bibitem{mizumoto}
Mizumoto, K. et al. (2020) Estimating the asymptomatic proportion of corona-virus disease 2019 (COVID-19) cases on board the Diamond Princess cruise ship, Yokohama, Japan, 2020. Euro Surveill. 25, pii=2000180. doi:10.2807/1560-7917.ES.2020.25.10.2000180

\bibitem{nishiura}
Nishiura, H. et al. (2020) Estimation of the asymptomatic ratio of novel coronavirus infections (COVID-19). Int. Journal of Infectious Diseases, doi:10.1016/j.ijid.2020.03.020 

\bibitem{lin}
Lin, M. et al. (2003) Association of HLA class I with severe acute respiratory syndrome coronavirus infection. BMC Med Genet. 4, 9

\bibitem{wang}
Wang, Z. et al. (2009) RNA-Seq: a revolutionary tool for transcriptomics. Nature reviews Genetics. 10, 57-63

\bibitem{warren}
Warren, R.L. et al. (2012) Derivation of HLA types from shotgun sequence datasets. Genome Med. 4, 95

\bibitem{brown}
Brown, S.D. et al. (2014) Neo-antigens predicted by tumor genome meta-analysis correlate with increased patient survival. Genome Res. 24, 743-750

\bibitem{wu}
Wu F. et al. (2020) Complete genome characterisation of a novel coronavirus associated with severe human respiratory disease in Wuhan, China. bioRxiv. doi:10.1101/2020.01.24.919183

\bibitem{sun}
Sun, Y. and Xi, Y. (2014) Association between HLA gene polymorphism and the genetic susceptibility of SARS infection. Book: HLA and associated important diseases. IntechOpen. 12, doi:10.5772/57561 

\bibitem{adamashvili}
Adamashvili, I. et al. (1997) Soluble HLA class I antigens in patients with type I diabetes and their family members. Human Immunol. 55, 176–83

\bibitem{noble}
Noble, J.A. et al. (2002) The HLA class I A locus affects susceptibility to type 1 diabetes. Human Immunol. 63, 657–664

\bibitem{nakanishi}
Nakanishi, K. and Inoko, H. (2006) Combination of HLA-A24, -DQA1*03, and -DR9 contributes to acute-onset and early complete beta-cell destruction in type 1 diabetes: longitudinal study of residual beta-cell function. Diabetes. 55, 1862–1868

\bibitem{kronenberg}
Kronenberg, D. et al. (2012) Circulating preproinsulin signal peptide-specific CD8 T cells restricted by the susceptibility molecule HLA-A24 are expanded at onset of type 1 diabetes and kill $\beta$-cells. Diabetes. 61, 1752–1759

\bibitem{guan}
Guan, W.J. et al. (2020) Comorbidity and its impact on 1590 patients with Covid-19 in China: A Nationwide Analysis. Eur Respir J. doi:10.1183/13993003.00547-2020

\bibitem{chu}
Chu, X. et al. (2018) Fine mapping MHC associations in Graves’ disease and its clinical subtypes in Han Chinese. J Med Genet. 55, 685–692

\bibitem{ollila}
Ollila, H.M. et al. (2015) HLA-DPB1 and HLA class I confer
risk of and protection from narcolepsy. Am J Hum Genet. 96, 136–146

\bibitem{kamatani}
Kamatani, Y. et al. (2009) A genome-wide association study identifies variants in the HLA-DP locus associated with chronic hepatitis B in Asians.
Nat. Genet. 41, 591-595

\bibitem{nguyen}
Nguyen, A. et al. (2020) Human leukocyte antigen susceptibility map for SARS-CoV-2.
medRxiv. doi:10.1101/2020.03.22.20040600

\bibitem{prachar}
Prachar, M. et al. (2020) COVID-19 vaccine candidates: prediction and validation of 174 SARS-CoV-2 epitopes.
bioRxiv. doi:10.1101/2020.03.20.000794

\end{thebibliography}

%

\end{document}